# Fluoride microresonators for mid-IR applications

IVAN S. GRUDININ,* KAMJOU MANSOUR, AND NAN YU

*Jet Propulsion Laboratory, California Institute of Technology, 4800 Oak Grove dr., Pasadena, California 91109, USA*
*Corresponding author: grudinin@jpl.nasa.gov*

**We study crystalline fluoride microresonators for mid-infrared applications. Whispering gallery mode resonators were fabricated with BaF$_2$, CaF$_2$ and MgF$_2$ crystals. The quality factors were measured at wavelengths of $1.56~\mu m$ and $4.58~\mu m$. The impacts of fabrication technique, impurities, multiphonon absorption and surface water are investigated. It is found that MgF$_2$ resonators have room temperature Q factor of $8.3 \times 10^6$ at wavelength of $4.58~\mu m$, limited by multiphonon absorption. © 2016 California Institute of Technology. Government sponsorship acknowledged.**

The number of applications of crystalline whispering gallery mode (WGM) micro-resonators is growing due to unique combination of record optical quality factors (Q), wide selection of material properties and small mode volumes [1]. The choice of materials suitable for optical applications in the UV or mid-IR is limited in comparison to the visible and near-IR wavelength ranges. Fluorides are unique as they not only have optical losses and Rayleigh scattering lower than that of fused silica [2], but also because of transparency in a broad wavelength range from 0.1 to 10 micrometers.

Optical frequency comb generation, an important application of microresonators, has been rapidly maturing into an applied technology [3, 4, 5]. As frequency combs are useful for spectroscopy in the mid-IR where spectral "fingerprints" of molecular species are located [6], it is attractive to develop mid-IR generators based on microresonators (microcombs). Several reports of mid-IR microcombs have been published recently [7, 8, 9]. The threshold of comb generation is inversely proportional to the Q factor squared [10], which makes Q one of the most important parameter in practical design of microcombs. As such, we have been working on achieving the highest possible values of Q factors in fluoride microresonators in the mid-IR. In particular for resonators made with MgF$_2$ – a convenient material for comb generation. Interestingly, in one recent report [9], generation of a comb is observed despite the fact that measured MgF$_2$ resonator Q is too low given the available laser pump power. This discrepancy was explained by a possibility that the real value of resonator Q was much higher than the value directly measured. Very high mid-IR Q factor values for MgF$_2$ resonators were also reported by another group, along with efficient frequency comb generation [8]. Given the intriguing development in this field and our own research effort in microcombs during the past few years, we set out to explore the potential of fluoride resonators for mid-IR applications. We experimentally measured the Q factors of MgF$_2$, CaF$_2$ and BaF$_2$ microresonators and investigated various relevant loss mechanisms affecting their performance.

We started by fabricating the resonators out of single-crystal fluoride windows in which the optical axis is coincident with the window axis (Z-cut). After the resonator blank is cut its edge is shaped with diamond-coated films and polished with glycol-based diamond suspensions. After polishing the resonators were cleaned as discussed below. Typical resonators are 3-7 mm in diameter and 1 mm thick with the sharp edge as shown in Fig. 1(a). The Q factors were first measured with a near-IR DFB fiber laser Koheras Adjustik emitting at $\lambda = 1561~nm$. Light was coupled into resonators with the couplers based on angle-polished single mode fibers [11]. In our initial set of near-IR measurements the resonators made from all three fluorides had similar values of resonance half-width at half-maximum (linewidth) of around 100 kHz under critical coupling conditions, corresponding to $Q \simeq 2 \times 10^9$. Critical coupling takes place when coupling loss equals intrinsic resonator loss. The measured values of around 100 kHz were likely limited by surface scattering as all three resonators were made with the same polishing and cleaning sequence. After near-IR characterization, the resonators were transferred into the mid-IR measurement setup. Clean ambient air environment in both setups and the transfer containers was maintained with HEPA filters having 99.99% efficiency for 0.3 µm particles.

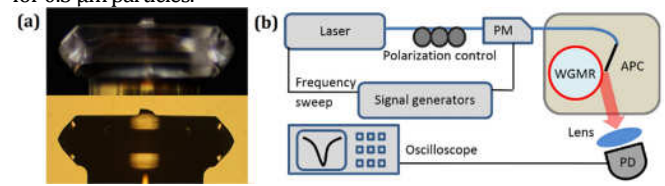

Fig. 1. (a) MgF$_2$ resonator under two different illumination conditions, diameter 3.1 mm. (b) Experimental setup schematics for 1561 nm laser. APC - angle-polished coupler, PD-photodetector, PM-phase modulator.

The mid-IR setup used a single mode DFB quantum cascade laser (QCL) emitting up to 70 mW near $\lambda = 4.58~\mu m$. The laser chip was integrated into a high heat load package along with thermoelectric modules for chip and housing temperature control. The laser beam was collimated with a high quality anti-reflection coated lens and sealed in our lab under dry air environment. The QCL frequency can be tuned or modulated by changing injection current. The FIO-4-4.5 isolator from Innovation Photonics was used to prevent the laser beam from reflecting back into the QCL. Careful alignment of the isolator was required to eliminate optical feedback caused by reflections from the isolator input window. Such feedback significantly affected the measured value of the resonator linewidth. Laser beam was expanded and then focused with a ZnSe lens onto a coupling prism. Mid-IR resonance linewidths in under-coupled configuration are summarized in Table 1.

**Table 1. Mid-IR measurement results and parameters**

| Resonator | Coupler | Linewidth | Quality factor |
|---|---|---|---|
| MgF$_2$ | CaF$_2$ | 8 MHz | $8.2 \times 10^6$ |
| CaF$_2$ | BaF$_2$ | 4 MHz | $1.6 \times 10^7$ |
| BaF$_2$ | Sapphire | <2MHz* | $> 3.2 \times 10^7$ |

*measurable linewidth is limited by laser noise*

The resonance linewidths were calibrated with the sideband modulation technique, where phase or frequency modulation is applied. This was achieved by a phase modulator in the near-IR setup and by direct QCL injection current modulation in the mid-IR setup. In the latter case, when QCL current is modulated at 25 MHz we could observe the sidebands that allowed us to calibrate the response of the laser frequency to current changes (blue trace in Fig. 2). After mid-IR Q value measurements each resonator was transferred back into the near-IR setup and its Q was confirmed to assure the absence of contamination during the mid-IR measurements. A typical measurement trace and its Lorentzian fit are given in Fig. 2, where calibration sidebands are also shown.

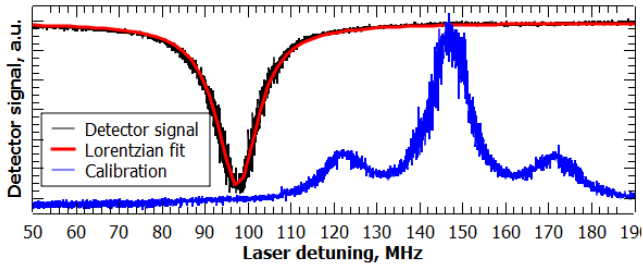

Fig.2. Linewidth measurement of a sample $MgF_2$ resonator at $\lambda = 4.58\ \mu m$ with a DFB QCL. Lorentzian fit linewidth is 11.6 MHz. Sidebands corresponding to QCL injection current modulation at a frequency of 25 MHz are also shown in blue.

The measured Q values presented in Table 1 qualitatively agree with the generally known mid-IR transparency of given fluorides. Yet, the Q values achieved in our lab in $MgF_2$ resonators corresponded to comb generation threshold well above the available mid-IR pump power of about 35 mW. Hence, we focused on investigations of limitations on Q factors in $MgF_2$ resonators. It is worth noting that recently published $MgF_2$ results [8] are a factor of 25 better than what we observed in mid-IR at nearly the same wavelength $\lambda = 4.58\ \mu m$ and about a factor of 10 better in near-IR. It is natural to expect that materials obtained from various vendors may have different optical losses. Moreover, we have observed a photorefractivity-like behavior with avoided crossing of modes during near-IR measurements. When laser frequency is scanned at 25 Hz over a 70 MHz window that contains a resonator mode as observed on the oscilloscope, one can see the appearance of a side mode with the frequency approximately 20 MHz below that of the main mode (on the red side). The side mode then moves across the main mode causing it to split in 2 or sometimes 3 modes in a way similar to avoided crossing. The secondary mode finally moves towards the blue side of the main mode. The whole process takes a few minutes when the pump power is 10 $\mu W$. Similar effects due to coupling of modes with different polarizations were also observed recently [12]. Since the observed behavior is somewhat similar to photorefractivity [13], it is reasonable to suggest that impurities causing photorefractivity might also be limiting the Q factor in $MgF_2$ resonators. It is also possible that the photorefractivity-like behavior could be caused by surface contamination as discussed below.

In order to estimate contribution from impurities we attempted an X-ray fluorescence (XRF) analysis with JPL's Horiba XGT-7200 XRF Microscope. However, most of the $MgF_2$ crystals are grown today from powders that have total impurity content below 100 ppm spread out over various elements, which is too low for XRF methods to work well. Mass spectroscopy methods reveal that typical crystalline blanks from a specific vendor have single digit ppm concentrations of Li, Ti, Cl, Al, Na, Ca, K, Si, Zr for $MgF_2$ samples and B, P, Fe, Cl, Si, S, Mg, Al, Na, K for $CaF_2$ samples. Dedicated analysis of specific samples would be required to identify any impurity contribution to optical loss. Moreover, spectroscopic data for many of these elements in specific crystalline hosts is often not available. The mass spectroscopy methods are destructive in nature and not suitable for testing commercial crystalline windows in a cost effective manner. Thus, the only viable method of estimating optical loss of crystalline windows at $\alpha < 10^{-6}\ cm^{-1}$ level in our lab remains the fabrication of WGM resonators and measurement of their Q values. Upper limit on optical loss is related to measured quality factor $Q = 2\pi n/\alpha\lambda$, where n is refractive index.

We fabricated eight $MgF_2$ resonators ranging from 3 to 7 mm in diameter using crystalline windows obtained from different vendors, including one used in a recent report [8]. When possible, the windows from the same vendor were manufactured from different ingots grown in different years. We found that all resonators had a critically coupled resonance linewidth of around 100 kHz in near-IR ($\lambda = 1.56\ \mu m$) and around 10 MHz in mid-IR ($\lambda = 4.58\ \mu m$). Such difference cannot be explained by surface scattering losses as they are inversely proportional to third power of wavelength [14]. Nevertheless, to reduce surface contamination we improved our cleaning and particle detection methods as follows. In our routine cleaning we use a stereoscopic Nikon MM-40 measuring microscope, which supports magnification of up to 500x and allows one to visually detect particles on the resonator surface at the diffraction limited resolution of around 0.5 μm. To improve particle detection further and to make resonator cleaning more reproducible we reflected 2 mW of collimated visible laser beam off the resonator surface. While reflection from highly polished crystalline surfaces does not create much of a background, one can still see a rather intense scattering from every particle on the resonator surface in the microscope field of view. Although this particle detection method is still subject to variations in human eye capabilities, the scattering intensity is not determined by the microscope resolution but only by the particle size. This allowed us to eliminate particles that could not be observed on the surfaces of resonators without using this method.

The difference between the near and mid IR Q values could also be explained by the presence of surface organic films that remain after cleaning. We use the purest, HPLC grade isopropanol and distilled water along with lint free paper wipes for cleaning. We found that moderate heating of the resonators after cleaning not only improves the near-IR Q factor but also eliminates the peculiar photorefractivity-like behavior described earlier in this paper. While comprehensive study of materials obtained from different vendors was beyond the scope of this work, we were able to improve the near-IR linewidth from 100 kHz to nearly 40 kHz for two different $MgF_2$ resonators by baking them for 1 hour at 250°C in dust-free ambient air. One of these two resonators was fabricated with the material used in [8]. Similar results were obtained for 10 hours baking at 120°C. We have also achieved good results with the UV-ozone cleaning method [15] using an 18 Watt mercury discharge lamp. However, heat treatment proved to be more effective. While 1 hour of UV cleaning improved the near-IR Q factor significantly, the photorefractivity-like behavior could not be eliminated. Usually, when we excite a WGM resonance in a cleaned resonator with a visible laser emitting near $\lambda = 0.65\ \mu m$, it is possible to see the belt of light on the surface caused by surface scattering (e.g. Fig. 2 in [16]). The belt is homogeneous if scattering is uniform or it can contain clearly visible scattering centers due to remaining particles or surface defects. In our case no particles or defects were present after the improved cleaning process described above. Interestingly, this belt could not be observed in resonators after baking. Importantly, we did not observe any Q improvement at mid-IR wavelength of $\lambda = 4.58\ \mu m$ in heat-treated $MgF_2$ resonators, which suggests that Q is limited by other mechanisms.

While our experiments with heat treatment of fluoride resonators provided encouraging results, a more dedicated study is needed to elucidate surface mechanism that affect optical Q. One possible explanation is the presence of thin organic film that forms on the surface during cleaning with solvents [17]. In fact, our baking experiments were motivated by the observations of residue that forms after the HPLC solvent is evaporated. On the other hand, it is known that ambient water vapor and carbon dioxide interact with the surface of fluoride crystals and form chemically and morphologically complex multiple-monolayer structures [18, 19]. While liquid water has strong absorption in the mid-IR range, the surface water layers have unique structure which is specific to particular crystals. Surface layers have characteristic mid-IR absorption bands that will shift and gradually disappear as the crystal is heated under vacuum [18, 19]. While evacuation under 400°C can be enough to eliminate water from CaF$_2$ and BaF$_2$, evacuation under 600-700°C is required to completely eliminate hydroxyl groups from MgF$_2$ [20]. Vacuum baking that would be required to eliminate all surface layers was beyond the scope of this work and will require a future study.

The fundamental limitation on the mid-IR quality factor of fluoride resonators is the multiphonon absorption [21, 22]. This mechanism is responsible for the mid-IR transparency cutoff of ionic solids. The absorption coefficient $\alpha(\omega, T)$, where T is temperature, decreases nearly exponentially with increasing frequency $\omega$. At room temperature and above there is little structure in $\alpha(\omega)$ for fluorides. Impurities can only increase $\alpha(\omega, T)$ above the multiphonon value at a given temperature. In the region of mid-IR cutoff no contribution from impurities was detected even for the crystal purity levels available in 1976 [21]. We have measured the mid-IR cutoff in our crystalline windows and determined the multiphonon absorption as follows. The transmittance of windows was directly measured with a Thermo Scientific nicolet iS50R FT-IR spectrometer. The windows were slightly tilted to eliminate interference errors due to reflections. The electric field of probing light was nearly orthogonal to optical crystalline axis. Room temperature transmittance for three windows of different thicknesses is shown in Fig. 3.

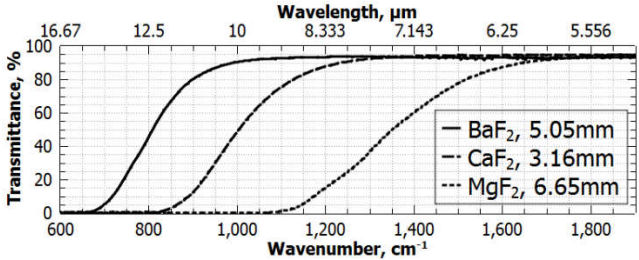

Fig. 3. Transmittance of three fluoride windows of different thickness in mid-IR as measured by the Fourier transform spectrometer.

We estimate absorption coefficient $\alpha$ at different frequencies from transmittance as follows. Single window transmittance T is given by $100\% \times I/I_0 = T_i^2 e^{-\alpha d}$, where $T_i = 4n/(n+1)^2$ is the theoretical single interface transmission coefficient, and $d$ is window thickness. Since exponential dependence is expected [21] we obtain $\log(\alpha) = \log\left(\frac{1}{d}\ln\left(\frac{16n^2}{0.01T(n+1)^4}\right)\right)$. Our window thickness measurement error is $\Delta d = 0.001\ cm$, and FT-IR spectrometer noise is $\Delta T = 0.22\%$. Thus we have the following measurement error: $\Delta \log(\alpha) = \frac{1}{d\ln(10)}\sqrt{(\Delta d)^2 + \left(\frac{1}{\alpha}\frac{\Delta T}{T}\right)^2}$. In these calculations we take dependence of refractive index on frequency into account with Sellmeier equations. These equations are available for wavelength up to $\lambda = 7\ \mu m$ for MgF$_2$ [23], to $\lambda = 12\ \mu m$ for CaF$_2$ and to $\lambda = 15\ \mu m$ for BaF$_2$ [24]. The results for $\log(\alpha)$ for all three fluorides are shown in Fig. 4. Measurement errors for MgF$_2$ are given as gray-colored band. The linear fit parameters are summarized in Table 2. In our data analysis we did not take multiple reflections inside each window into account [22] when computing absorption coefficients as the expected correction is negligible and our results match well with the earlier measurements.

Table 2. Fit parameters for $\log(\alpha[cm^{-1}]) = a - b/(\lambda[cm])$

| Window material | a | b | Fitting range |
|---|---|---|---|
| MgF$_2$ | 4.3 | $3.19 \times 10^{-3}$ | 1200-2000 |
| CaF$_2$ | 4.88 | $4.52 \times 10^{-3}$ | 830-1100 |
| BaF$_2$ | 5.04 | $6.1 \times 10^{-3}$ | 680-1100 |

The linear fit for MgF$_2$ was extended beyond the fitting range to $\omega = 2250\ cm^{-1}$ as a guideline. While we could not directly measure the absorption up to $\omega = 2250\ cm^{-1}$, there is currently no reason to expect any significant deviation from the exponential law [25]. The *MgF$_2$ 2003* data shown in Fig. 4 was taken from Corning material specification and could not be traced to any published measurement. On the other hand, there is a good agreement between our measurement and earlier data. The discrepancy between our data and recently reported results [8] is yet to be explained.

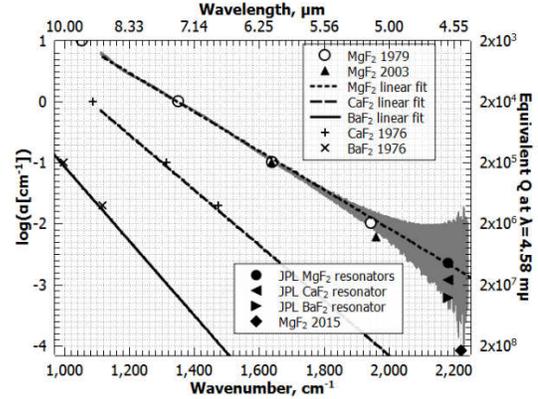

Fig. 4. Mid-IR absorption of three fluorides and experimental results for microresonators. The gray band represents experimental measurements for MgF$_2$ with error bars. Experimental data from 1976 [21], 1979 [22] and 2003 [26] match our measurement results well. Recent 2015 data for MgF$_2$ [8] is also shown.

It is worth noting that the difference between our experimental transmittance for MgF$_2$ and theoretical $T_i^2$ is less than 0.05% in the region of wavenumbers $2400 - 2800\ cm^{-1}$.

We have also measured the Q of an MgF$_2$ resonator in the mid-IR ($\lambda = 4.58\ \mu m$) at a temperature of approximately 140°C. We found that the Q value dropped to about $5.5 \times 10^6$ from the room temperature value of $8 \times 10^6$. We then measured the transmittance of our MgF$_2$ window both at room temperature and at around 140°C and found that the absorption coefficient increases by around 20% for any wavelength in the multiphonon-limited spectral region. This preliminary observation is qualitatively consistent with the temperature dependence of multiphonon absorption and its role in limiting the Q value of MgF$_2$ resonators in the mid-IR wavelength region. Along these lines, experimental data [22] suggests that it is possible to significantly reduce the multiphonon loss by lowering crystal temperature. Finally, using our linear fit data, we plot multiphonon-limited Q factors of fluoride resonators at room temperature in the mid-IR region as shown in fig. 5. The dependence of refractive indices on wavelength was accounted for.

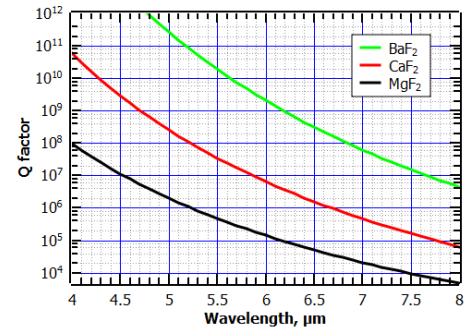

Fig. 5. Q factor of fluoride resonators as limited by multiphonon absorption at room temperature.

In conclusion, we have described a cleaning and heat treatment procedure that can improve optical Q of fluoride resonators. We observed no improvement of mid-IR Q for MgF$_2$ resonators beyond the multiphonon absorption limited value of $8.3 \times 10^6$ at $\lambda = 4.58\ \mu m$. Importantly, the Q values reported here for CaF$_2$ and BaF$_2$ resonators were obtained without the baking procedure. Thus, the ultimate experimental mid-IR Q factors of resonators made with these and similar materials may be much higher [27, 28] and will be reported elsewhere. A more detailed study is also required to quantify the influence of surface layers of water and other chemicals on optical loss of fluoride resonators. While MgF$_2$ is attractive for such applications as mid-IR frequency comb generation and ringdown spectroscopy due to its ability to be thermally self-locked to the laser frequency, other fluorides represent attractive alternatives due to lower multiphonon losses.

**Acknowledgment**. The research was carried out at the Jet Propulsion Laboratory, California Institute of Technology, under a contract with the National Aeronautics and Space Administration. We thank OEWaves for sharing MgF$_2$ samples and Q measurement results, helpful discussions and clarifications.

## References

[1] V. S. Ilchenko and A. B. Matsko, "Optical Resonators With Whispering-Gallery Modes - Part II: Applications," *IEEE J. Sel. Top.Quant. Electron.,* vol. 12, no. 1, pp. 15-32, 2006.

[2] S. Logunov and S. Kuchinsky, "Experimental and theoretical study of bulk light scattering in CaF2 monocrystals," *J. Appl. Phys. ,* vol. 98, p. 053501, 2005.

[3] V. Brasch, M. Geiselmann, T. Herr, G. Lihachev, M. H. P. Pfeiffer, M. L. Gorodetsky and T. J. Kippenberg, "Photonic chip-based optical frequency comb usning soliton Cherenkov radiation," *Science,* vol. science.aad4811, p. 10.1126, 2015.

[4] W. Liang, D. Eliyahu, V. S. Ilchenko, A. A. Savchenkov, A. B. Matsko, D. Seidel and L. Maleki, "High spectral purity Kerr frequency comb radio frequency photonic oscillator," *Nature Commun.,* vol. 6, p. 7957, 2015.

[5] S. A. Miller, Y. Okawachi, S. Ramelow, K. Luke, A. Dutt, A. Farsi, A. L. Gaeta and M. Lipson, "Tunable frequency combs based on dual microring resonators," *Opt. Express,* vol. 23, no. 16, pp. 21527-21540, 2015.

[6] A. Schliesser, N. Picque and T. W. Haensch, "Mid-infrared frequency combs," *Nature Photon.,* vol. 6, pp. 440-449, 2012.

[7] K. Luke, Y. Okawachi, M. R. Lamont, A. L. Gaeta and M. Lipson, "Broadband mid-infrared frequency comb generation in a Si3N4 microresonator," *Opt. Lett.,* vol. 40, no. 21, pp. 4823-4826, 2015.

[8] A. A. Savchenkov, V. S. Ilchenko, T. F. Di, P. M. Belden, W. T. Lotshaw, A. B. Matsko and L. Maleki, "Generation of Kerr combs centered at 4.5 mu m in crystalline microresonators pumped with quantum-cascade lasers," *Opt. Lett.,* vol. 40, no. 15, pp. 3468-3471, 2015.

[9] C. Lecaplain, C. Javerzac-Galy, E. Lucas, J. D. Jost and T. J. Kippenberg, "Quantum cascade laser Kerr frequency comb," *arxiv:1506.00626,* 2015.

[10] A. B. Matsko, A. A. Savchenkov, D. Strekalov, V. S. Ilchenko and L. Maleki, "Optical hyperparametric oscillations in a whispering-gallery-mode resonator: Threshold and phase diffusion," *Phys. Rev. A,* p. 033804, 2005.

[11] V. S. Ilchenko, S. X. Yao and L. Maleki, "Pigtailing the high-Q microsphere cavity: a simple fiber coupler for optical whispering-gallery modes," *Opt. Lett.,* vol. 24, no. 11, pp. 723-725, 1999.

[12] W. Weng and A. N. Luiten, "Mode-interactions and polarization conversion in a crystalline microresonator," *Optics Letters,* vol. 40, no. 23, pp. 5431-5434, 2015.

[13] A. A. Savchenkov, A. B. Matsko, D. Strekalov, V. S. Ilchenko and L. Maleki, "Photorefractive effects in magnesium doped lithium niobate whispering gallery mode resonators," *Appl. Phys. Lett.,* vol. 88, p. 241909, 2006.

[14] M. L. Gorodetsky, A. D. Pryamikov and V. S. Ilchenko, "Rayleigh scattering in high-Q microspheres," *J. Opt. Soc. Am. B,* vol. 17, no. 6, pp. 1051-1057, 2000.

[15] J. R. Vig, "UV/ozone cleaning of surfaces," *J. Vac. Sci. Technol.,* vol. A3, no. 3, pp. 1027-1034, 1985.

[16] M. L. Gorodetsky and I. S. Grudinin, "Fundamental thermal fluctuations in microspheres," *J. Opt. Soc. Am. B,* vol. 21, no. 4, pp. 697-705, 2004.

[17] R. E. Robinson and et al., "Removing Surface Contaminants from Silicon Wafers to Facilitate EUV Optical Characterization," in *47th Annual Technical Conference Proceedings*, Dallas, 2004.

[18] G. E. Ewing, "Ambient thin film water on insulator surfaces," *Chem. Rev. ,* vol. 106, pp. 1511-1526, 2006.

[19] T. S. Minakova and I. A. Ekimova, Фториды и оксиды щелочноземельных металлов и магния. Поверхностные свойства., Томск: Издательский дом ТГУ, 2014.

[20] P. B. Barraclough and P. G. Hall, "Adsorption of water vapour by magnesium fluoride," *J. Chem. Soc., Faraday Trans. 1,* vol. 72, pp. 610-618, 1976.

[21] H. G. Lipson, B. Bendow, N. E. Massa and S. S. Mitra, "Multiphonon infrared absorption in the transparent regime of alkaline-earth fluorides," *Phys. Rev. B,* vol. 13, no. 6, pp. 2614-2619, 1976.

[22] B. Bendow, H. G. Lipson and S. S. Mitra, "Multiphonon infrared absorption in highly transparent MgF2," *Phys. Rev. B,* vol. 20, no. 4, pp. 1747-1749, 1979.

[23] M. J. Dodge, "Refractive properties of magnesium fluoride," *Appl. Opt.,* vol. 23, pp. 1980-1985, 1984.

[24] H. H. Li, "Refractive index of alkaline earth halides and its wavelength and temperature derivatives," *J. Phys. Chem. Ref. Data,* vol. 9, pp. 161-289, 1980.

[25] P. A. Miles, "Temperature dependence of multiphonon absorption in zinc selenide," *App. Opt.,* vol. 16, no. 11, pp. 2891-2896, 1977.

[26] Corning, "Magnesium fluoride product data," 2003.

[27] M. diciliani de Cumis, S. Borri, G. Insero, I. Galli, A. Savchenkov, D. Eliyahu, V. Ilchenko, N. Akikusa, A. Matsko, L. Maleki and P. De Natale, "Microcavity-Stabilized Quantum Cascade Laser," *Laser Photonics Rev.,* vol. 10, no. 1, pp. 153-157, 2016.

[28] B. Way, R. K. Jain and M. Hossein-Zadeh, "High-Q microresonators for mid-IR light sources and molecular sensors," *Opt. Lett.,* vol. 37, no. 21, pp. 4389-4391, 2012.

[29] W. Kern, Handbook of semiconductor wafer cleaning technology, Park Ridge: Noyes Publications, 1993.